\documentclass[10pt, conference, letterpaper]{IEEEtran}
\IEEEoverridecommandlockouts
\usepackage{algpseudocode}
\usepackage[linesnumbered,ruled]{algorithm2e}
\usepackage{algorithmicx}
\usepackage{stfloats}
\usepackage[caption=false,font=footnotesize]{subfig}

\usepackage{cite}
\usepackage{amsmath,amssymb,amsfonts}
\usepackage{graphicx}
\usepackage{textcomp}
\usepackage{xcolor}
\def\BibTeX{{\rm B\kern-.05em{\sc i\kern-.025em b}\kern-.08em
    T\kern-.1667em\lower.7ex\hbox{E}\kern-.125emX}}
\begin{document}

\title{Online Caching and Coding at the WiFi Edge: Gains and Tradeoffs}

\author
{
\IEEEauthorblockN{Lalhruaizela Chhangte
}
\IEEEauthorblockA{IITB-Monash Research Academy \\ Mumbai, India \\ 154074004@iitb.ac.in}
\and
\IEEEauthorblockN{Emanuele Viterbo 
}
\IEEEauthorblockA{Monash University \\ Clayton, VIC, Australia \\ emanuele.viterbo@monash.edu}
\and
\IEEEauthorblockN{D Manjunath
}
\IEEEauthorblockA{IIT Bombay \\ Mumbai, India \\ dmanju@ee.iitb.ac.in}
\and
\IEEEauthorblockN{Nikhil Karamchandani
}
\IEEEauthorblockA{IIT Bombay \\ Mumbai, India \\ nikhilk@ee.iitb.ac.in}
}

\maketitle

\begin{abstract}

Video content delivery at the wireless edge continues to be challenged by insufficient bandwidth and highly dynamic user behavior which affects both effective throughput and latency. Caching at the network edge and coded transmissions have been found to improve user performance of video content delivery. The cache at the wireless edge stations (BSs, APs) and at the users' end devices can be populated by pre-caching content or by using online caching policies. In this paper, we propose a system where content is cached at the user of a WiFi network via online caching policies, and coded delivery is employed by the WiFi AP to deliver the requested content to the user population. The content of the cache at the user serves as side information for index coding. We also propose the LFU-Index cache replacement policy at the user that demonstrably improves index coding opportunities at the WiFi AP for the proposed system. Through an extensive simulation study, we determine the gains achieved by caching and index by coding. Next, we analyze the tradeoffs between them in terms of data transmitted, latency, and throughput for different content request behaviors from the users. We also show that the proposed cache replacement policy performs better than traditional cache replacement policies like LRU and  LFU.

\end{abstract}


\section{Introduction}

Video constitutes more than 70\% of the total IP traffic today, and is expected to reach 82\% by 2020\cite{cisco}. Video content delivery at the wireless edge continues to be challenged by insufficient bandwidth and highly dynamic user behavior. This challenge will be further exacerbated because mobile and wireless devices are expected to account for two-thirds of the total IP traffic by 2020\cite{cisco}. Two mechanisms that can be used to improve user performance of video delivery are: (i)~caching of video content at the wireless edge nodes---base stations (BSs), access points (APs), users and (ii)~coded delivery techniques that multicast, or broadcast, a single stream to serve multiple users simultaneously. 

Caching of content at the network edge has been found to be useful to improve content delivery by several recent investigations \cite{femto, livingedge, wirelessdtod, energy, pushed}. However, most of these studies considered pre-caching of contents based on the predicted users' demand profile \cite{femto, livingedge, vidaware, informed}. On the other hand, only a few studies considered online caching, where the contents to be cached are decided by cache replacement policies after they are requested \cite{vidaware, onlinecoded}. Online caching is expected to be increasingly important because popularity profiles of content is expected to change on a faster timescale. 

A second mechanism that can help improve user performance of video content delivery is coded delivery. Coded delivery techniques \cite{fundamental, decentralized, iscod} are information-theoretic approaches that use side information cached at the client\footnote{Throughout the paper, we use client and user interchangeably to denote user terminal.} side so that the server can broadcast index-coded information simultaneously to multiple clients. This broadcast information is simultaneously received by multiple clients who use the side information in their local caches to decode the information that they need. The side information, which are the contents cached at the users can be pre-cached \cite{fundamental, decentralized} or updated via online methods \cite{onlinecoded, onlinerandom}. Also, the works on online coded delivery techniques considered the long-term (weeks, months) gains for evaluating the performance of coding. However, user behavior in a wireless network is highly dynamic, and the clients connected to a single wireless edge station (BS, AP) can change on faster timescales. 

Our interest in this work is to assess how frequently index coding opportunities arise in a WiFi network over short timescales and to estimate the advantages of caching and index coding on throughput and latency as a function of content requests behavior from the users. The following are the specific contributions of this paper.

\begin{itemize}
    \item From our experience of building the Wi-Cache system \cite{wicache,wicache2}, we propose a system model to exploit possible gains of online caching and index coding. 
    \item For this system, we propose the LFU-Index cache replacement policy at the user and a coding heuristic at the wireless edge station, e.g., WiFi AP. 
    \item From an extensive simulation study, we determine gains and tradeoffs from caching and index coding using metrics like data transmitted, latency and throughput.  
\end{itemize}

The rest of the paper is organized as follows. In the next section we present an index coding problem instance. Section~\ref{sec:general} describes a general video content distribution system. Section~\ref{sec:proposed} contains the detailed proposed system model. Section~\ref{sec:simulation} describes the simulation settings while the key results are provided in Section \ref{sec:metrics}.

\section{Index coding}
\label{sec:index_coding}
\begin{figure}[ht]
  \centering
  \includegraphics[scale=0.5]{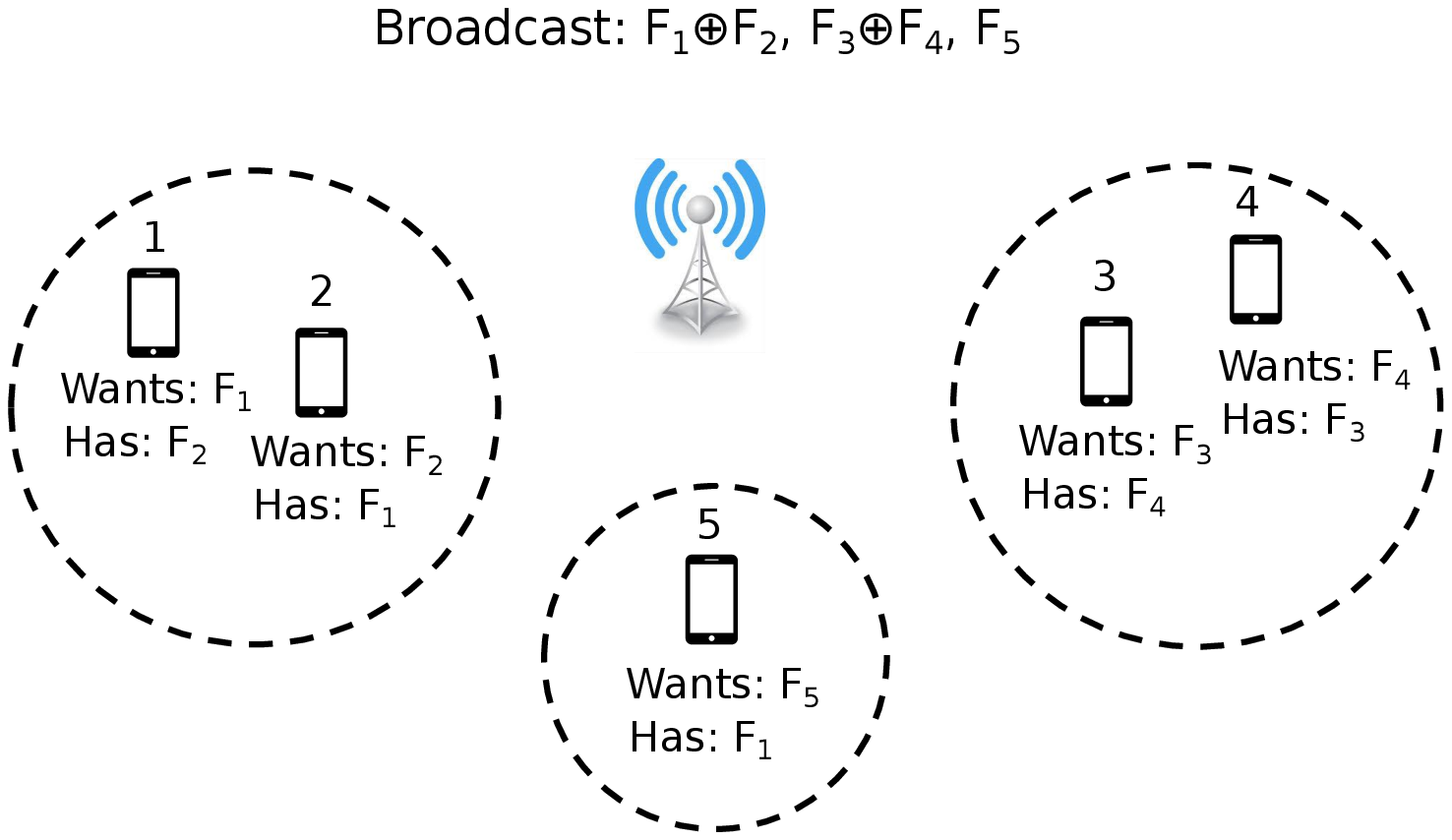}
  \caption{An index-coding problem instance}
  \label{fig:Index coding problem instance}
  \vspace{-4mm}
\end{figure}

Fig.~\ref{fig:Index coding problem instance} shows an index coding problem
instance with five wireless clients and a server. The server has the
set of files $\{F_{1},\ldots,F_{5}\}$, and each client $i$ wants file
$F_i$ from the server, denoted by the set $Wants$. Further, each
client has cached a copy of the files previously received from the
server, denoted by the set $Has.$ When the sets $Wants$ and $Has$ of
each client is revealed to the server, the server constructs a set of
coded files, and broadcasts the coded stream that is received by all
the clients. Each client then decodes the coded file to retrieve the
file that it wants. Coding and decoding is by simple $XOR$ operations.
In the example of Fig.~\ref{fig:Index coding problem instance}, after
receiving the sets $Wants$ and $Has$ of each of the clients, the
server broadcasts: $F_{1} \bigoplus F_{2}, F_{3} \bigoplus F_{4},
F_{5}$. Upon receiving the broadcast, clients~1 and~2 can retrieve
$F_{1}$ and $F_{2}$ from $F_{1} \bigoplus F_{2}$ by performing $XOR$
with $F_{2}$ and $F_{1}$ respectively. Similarly, clients 3 and 4 can
retrieve $F_{3}$ and $F_{4}$ from $F_{3} \bigoplus F_{4}$
respectively, and client 5 can receive $F_{5}$ as it is. Therefore,
instead of transmitting five different files, the server transmits
only 3 files thus reducing the number of bits transmitted over the
broadcast channel.

\section{Video content distribution system}
\label{sec:general}
\begin{figure}
  \centering
  \includegraphics[scale=0.6]{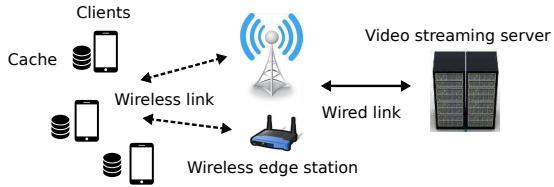}
  \caption{Video content distribution system}
  \label{fig:system_model}
  \vspace{-4mm}
\end{figure}

We consider a system consisting of a wireless edge station, which could be a WiFi access point (AP) or a cellular base station (BS) connected to a video streaming server through a wired link as shown in Fig. \ref{fig:system_model}. The video streaming server stores $N$ video files denoted by $F := \{f_{1},\ldots,f_{N}\}$, which are split into smaller video segments of fixed size playback duration $P$. $K$ clients denoted by $C := \{c_{1}$,\ldots,$c_{K}\}$ are connected to the wireless edge station through the wireless link of bandwidth $B_{w}$. Each client is equipped with a cache of size $M$ bits to cache the video segments requested from the server. 

\section{System model}
\label{sec:proposed}
We propose the following system model to study the performance of caching and index coding, as well as the tradeoffs between them. This model is based on the Wi-Cache wireless edge caching system that we have developed. The system architecture and the implementation details are available in \cite{wicache, wicache2}.

\subsection{File popularity model}
\label{subsec:file_pop_model}
Let $P^{(i)} := \{P_{f_{1}}^{(i)}, P_{f_{2}}^{(i)},\ldots, P_{f_{N}}^{(i)}\}$ be the file popularity distribution of the video files $f_{1}, f_{2},\ldots, f_{N}$ of a client $c_{i}$. In the proposed model, initially we assign the same file popularity distribution for all the clients, which is then updated by each client based on its own request.

\textit{Initial distribution:} We consider MZipf distribution to model the initial file popularity distribution for each client, an assumption supported by a recent measurement study on large-scale content distribution systems \cite{bbc}. Based on MZipf, the probability that users want to request a video file $f_{i}$, i.e., request probability, is denoted by

\[
P_{f_{i}} = \frac{(i+q)^{-\gamma}}{\sum_{j=1}^{N}(j+q)^{-\gamma}}, \text{$i$ = 1,2,\ldots,$N$}
\]
where $N$ is the total number of video files in the server, $\gamma$ is the Zipf factor, and $q$ is the plateau factor. 

\textit{Update:} A client $c_{i}$ updates its own distribution every time it requests a file based on a rewatch factor, denoted by $\alpha$, with $\alpha \in [0,1]$. If a client $c_{i}$ request a video file $f_{j}$, the new probabilities are calculated using the following

\begin{align*}
P_{f_{k}}^{(i)} &= P_{f_{k}}^{(i)} +\!\! \left(\! \frac{P_{f_{k}}^{(i)}}{1\!-\!P_{f_{j}}^{(i)}} \cdot (P_{f_{j}}^{(i)} - \alpha \cdot P_{f_{j}}^{(i)} )\! \right)\!,
\forall  f_{k} \in F \setminus \{f_{j}\} \\
P_{f_{j}}^{(i)} &= \alpha \cdot P_{f_{j}}^{(i)}
\end{align*}
That is, the request probability of the video file $f_{j}$ is scaled by $\alpha$, and the reduction in probability of the file $f_j$, $(P_{f_{j}}^{(i)} - \alpha \cdot P_{f_{j}}^{(i)})$,
is distributed across the other files proportionally to their current request probabilities.

\textit{Special cases:}
\begin{itemize}
    \item \textit{$\alpha=0$}: In this case, the probabilities of the requested files are updated to $0$, and therefore the clients do not rewatch video files.
    \item \textit{$\alpha=1$}: In this case, there is no change in the request probabilities, and therefore the clients maintain the same initial distribution independent of their requests.
\end{itemize}

\subsection{Video streaming and service model}
\label{sub:stream_model}
Streaming a video file $f_{i}$ which is split into $S$ number of segments involves requesting the set of video segments $f_{i}^{1}, f_{i}^{2},\ldots,f_{i}^{S}$ in sequential order, and one segment at a time. We consider that when a client initially joins the network, or finished streaming a video file, it waits for a random amount of time, $T_{w}$, before requesting a video file. We model $T_{w}$ as an exponential random variable with mean $1 \backslash \lambda_{S}$. 

\begin{figure}[ht]
  \centering
  \includegraphics[scale=0.5]{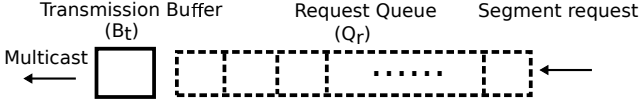}
  \caption{Service model}
  \label{fig:edge}
\end{figure}

We consider a system where the clients send their cache content information, i.e., the set of video segments present in their caches to the wireless edge station node every time they establish a new connection. Also, we require that the wireless edge station node keep track of the cache content information of a client as long as the client is connected to it. Once a client is connected to the wireless edge station, it selects a file based on the file popularity model described in Sec. \ref{subsec:file_pop_model}, and streams the video file. In order to handle the video segment requests that arrive asynchronously at the wireless edge station, we consider a transmission buffer $B_{t}$, and a request queue $Q_{r}$ as illustrated in Fig. \ref{fig:edge}.

When a segment request $r_{i}$ from a client $c_{i}$ arrives, it is placed at the end of $Q_{r}$. The video segment currently being transmitted is in $B_{t}$. As soon as the current transmission is finished, the video segment request at the front of $Q_{r}$ is placed in $B_{t}$, and the transmission of the video segment begins. In the system model, we consider that the transmission is a multicast, and the clients process the multicast segments which are only designated for them. The multicast segments can be either index-coded or non-index-coded.  

In order to serve the clients' requests, the wireless edge station fetches the content of the requested video segments from the video streaming server. Meanwhile, it also looks for index coding opportunities among the received segment requests. 

{\em Index coding model: }When index coding is enabled at the wireless edge station, for every segment request that arrives, the wireless edge station checks whether it can be index-coded with any of the segment requests currently in $Q_{r}$. For a segment request $r_{i}$ from client $c_{i}$, we define $H(r_{i})$ and $W(r_{i})$, where $H(r_{i})$ is the set of video segments currently cached by $c_{i}$, and $W(r_{i})$ is the set of video segments requested by $c_{i}$. Requests $r_{i}$ from $c_{i}$, and $r_{j}$ from $c_{j}$ can be index-coded if

\[
W(r_{i}) \subset H(r_{j}) \text{ and } W(r_{j}) \subset H(r_{i})
\]
and we represent the index-coded segment request as $r_{i,j}$. For the index-coded segment request $r_{i,j}$ which consists of requests from $c_{i}$ and $c_{j}$, $H(r_{i,j})$ = $H(r_{i}) \cap H(r_{j})$, and $W(r_{i,j})$ = $W(r_{i}) \cup W(r_{j})$. Given $\{W(r_{i}),H(r_{i})\}_{i=1,\cdots,K},$ finding an optimal strategy that maximizes the index coding is NP-hard \cite{indexalgorithms}; luckily many low complexity heuristic algorithms are known to perform well \cite{indexalgorithms}.

We also propose a heuristic to maximize the index coding performed at the wireless edge station. Algorithm \ref{algo:segment_coding} uses the proposed heuristic. In order to describe the heuristic, we first define the degree of freedom (DOF)\cite{iscod} and degreee of effort (DOE). Let us consider that a request $r_{j}$ from $c_{j}$ currently resides in $Q_{r}$, we call $|H(r_{j})|$ the DOF of $r_{j}$, and $|W(r_{j})|$ the DOE of $r_{j}$. DOF represents the potential number of other segment requests which can be served by the cache content of $c_{j}$ through index coding. DOE represents the number of segments that have to be present in the cache of any other client to form index coded segment with $r_{j}$. The heuristic employed by Algorithm \ref{algo:segment_coding} minimizes the loss of DOF, if there are multiple index coding opportunities for an incoming request, and it minimizes the propagation of DOE, if the loss of DOF is the same. 


\begin{algorithm}[tbh]
  \caption{Index coding}
  \label{algo:segment_coding}
  \SetAlgoLined
  $r_{i}$: segment request just arrived\;
  $Q_{r}$: Request queue\; 
  $sel\_req$: selected request\;
  $sel\_DOF$: $DOF$ of selected request\;
  $sel\_DOE$: $DOE$ of selected request\;
  $cur\_DOF$ = 0; $cur\_DOE$ = $\infty$\; 
  \While{end of $Q_{r}$ is not reached}{
    pick a segment request $r_{j}$ from the front of $Q_{r}$\;
    \uIf{($W(r_{i}) \subset H(r_{j}))$ and ($W(r_{j}) \subset H(r_{i})$)}{
      \uIf{$(DOF(r_{i,j}) > cur\_DOF)$}{
        $sel\_req$ = $r_{j}$; $sel\_DOF$ = $|H(r_{i,j})|$; $sel\_DOE$ = $|W(r_{i,j})|$;
      }
      \uElseIf{$(DOF(r_{i,j}) == cur\_DOF)$}{
        \uIf{($DOE(r_{i,j}) < cur\_DOE$)}{
          $sel\_req$ = $r_{j}$; $sel\_DOF$ = $|H(r_{i,j})|$; $sel\_DOE$ = $|W(r_{i,j})|$;
        }
      }
    }
  }
  \uIf{($sel\_DOF > 0$)}{
    $r_{sel\_req, r_{i}}$ = Code $sel\_req$ and $r_{i}$\;
    update $DOF$ and $DOE$ of $r_{sel\_req, r_{i}}$ 
  }
  \Else{
    put $r_{i}$ at the end of $Q_{r}$
  }
\end{algorithm}
  \vspace{-5mm}
\subsection{Caching model and cache replacement policies}
\label{subsec:cache_model}
We consider an online caching model at the clients, i.e., a client caches the video segment requested if there is enough space in its cache. If there is not enough space in its cache, it evicts one or more segments from its cache to accommodate the new segment based on a certain cache replacement policy. Moreover, when segment requests are found in the cache, i.e., there is a cache hit, the requests are served from the cache directly.

Traditional cache replacement policies such as least recently used (LRU)\cite{lru} and least frequently used (LFU)\cite{lfu} use simple heuristics to improve the cache hit count, i.e., the number of segment requests which are found in the cache. We consider the following model to describe the cache replacement policies used in the system model---LFU, LRU, Belady, LFU-Index. 

Let $S_{i} := \{s_{1}, s_{2}, ..\}$ be the set of all the segments present in the cache of a client $c_{i}$, and for each segment $s_{j}$ present in the cache, we define a 6-tuple representing the segment
\[<s_{j}, \beta_{s_{j}}^{l}, \Gamma_{s_{j}}^{l}, \beta_{s_{j}}^{g}, \Gamma_{s_{j}}^{g}, \tau_{s_{j}}>,\]
where $s_{j}$ is the segment name, $\beta_{s_{j}}^{l}$ is the last time $s_{j}$ was requested by $c_{i}$, $\Gamma_{s_{j}}^{l}$ is the request count of $s_{j}$ by $c_{i}$, $\beta_{s_{j}}^{g}$ is the last time $s_{j}$ was requested by any client in $C$, $\Gamma_{s_{j}}^{g}$ is the sum of the request count of $s_{j}$ across all the clients in $C$, and $\tau_{s_{j}}$ is the number of clients which cache $s_{j}$.  

\subsubsection{LRU}
\label{subsubsec:lru}
LRU evicts the segment with the least $\beta_{s_{j}}^{g}$ to accommodate the new segment.

\subsubsection{LFU}
\label{subsubsec:lfu}
LFU evicts the segment with the least $\Gamma_{s_{j}}^{g}$ to accommodate the new segment.

\subsubsection{Belady}
\label{subsubsec:belady}
Belady\cite{belady} is the optimal cache replacement policy in terms of cache hit count. It chooses to evict a segment in the cache which will be requested furthest in the future. However, it requires knowledge of the future and is thus not achievable in practice.

We also propose LFU-Index, a cache replacement policy which extends LFU by improving the index coding opportunities in addition to the cache hit count. In order for index coding to be possible between two clients, the request of each client has to be present in the cache of the other client. Therefore, index coding opportunities arise when the cache content of the clients in the network are different, and the clients request different contents. The heuristic employed by LFU-Index improves index coding opportunities by maximizing the difference in the content of caches across the clients, which is shown in Algorithm \ref{algo:lfu_index}. From the set of segments with minimum request count, i.e., $\Gamma^{g}=min\_\Gamma^{g}$, a client evicts the segment which is cached the most by the other clients, i.e., $\tau=max\_\tau$. Also, if there are multiple segments with $\tau=max\_\tau$, the segment with the least $\beta^{l}$ is evicted.

\begin{algorithm}[tbh]
  \caption{LFU-index}
  \label{algo:lfu_index}
  \SetAlgoLined
  $s$: new segment just received by client $c_{i}$\;
  $min\_\Gamma^{g}$: initialize with minimum $\Gamma_{s_{j}}^{g}$ from $S_{i}$\;
  $max\_\tau$ = 0\;
  $min\_\beta^{l} = \infty$\;
  $p\_s$: place holder for segment to be evicted\;
 \For{each segment $<s_{j}, \beta_{s_{j}}^{l}, \Gamma_{s_{j}}^{l}, \beta_{s_{j}}^{g}, \Gamma_{s_{j}}^{g}, \tau_{s_{j}}>$ in $S_{i}$}{
    \uIf{$\Gamma_{s_{j}}^{g} == min\_\Gamma^{g}$}{
      \uIf{$\tau_{s_{j}} > max\_\tau$}{
        $p\_s = s_{j}$\;
        $max\_\tau = \tau_{s_{j}}$\;
      }
      \uElseIf{$\tau_{s_{j}} == max\_\tau$}{
        \uIf{$\beta_{s_{j}}^{l} < min\_\beta^{l}$}{
          $p\_s = s_{j}$\;
          $min\_\beta^{l} = \beta_{s_{j}}^{l}$\;
        }
      }
    }
  }
    
  Evict $p\_s$ from $S_{i}$\;
  Add $s$ to $S_{i}$

\end{algorithm}

\vspace{-5mm}
\section{Simulation settings}
\label{sec:simulation}
\begin{figure}
  \centering
  \includegraphics[scale=0.4]{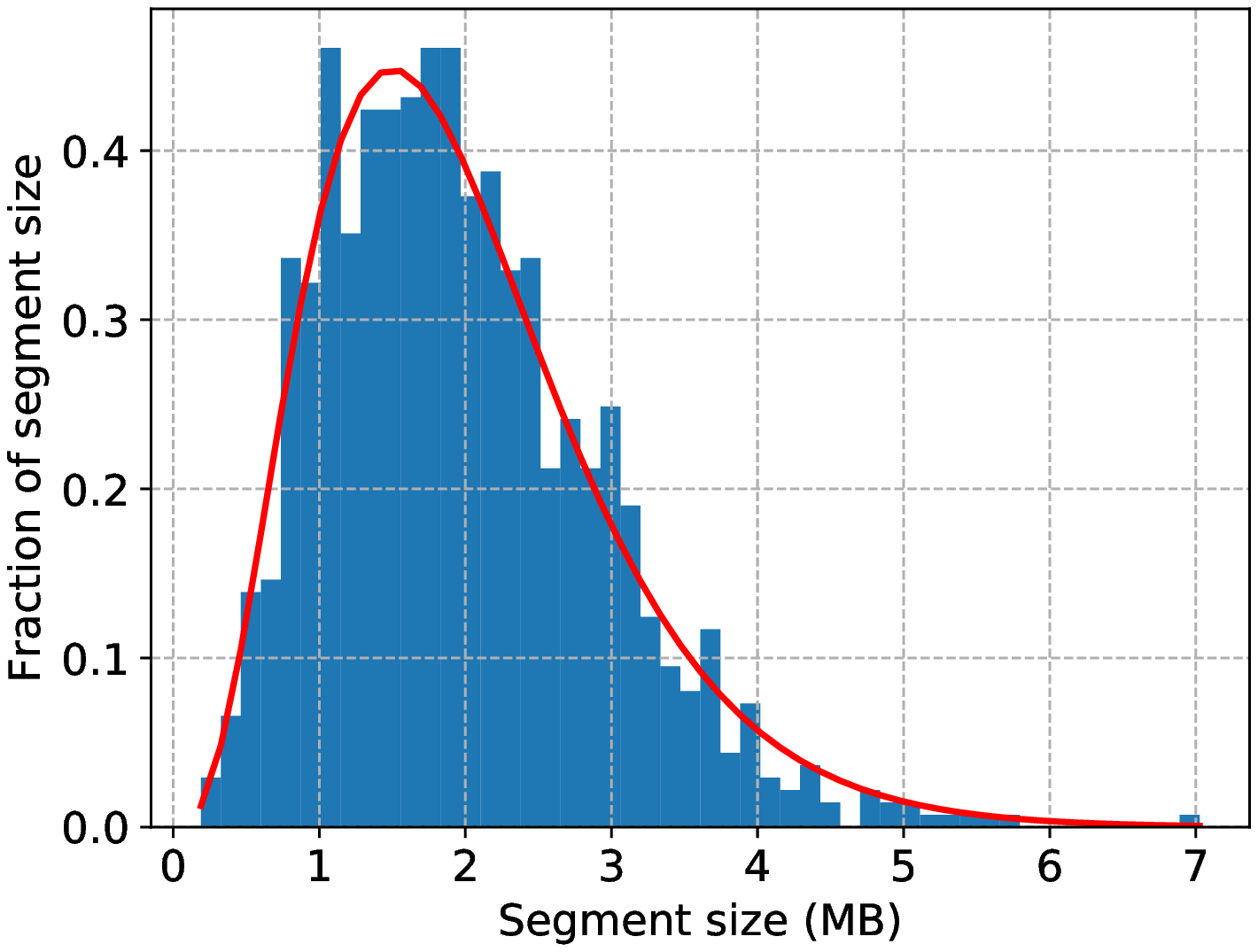}
  \caption{Segment size distribution}
  \label{fig:segment_dist}
  \centering
  \includegraphics[scale=0.4]{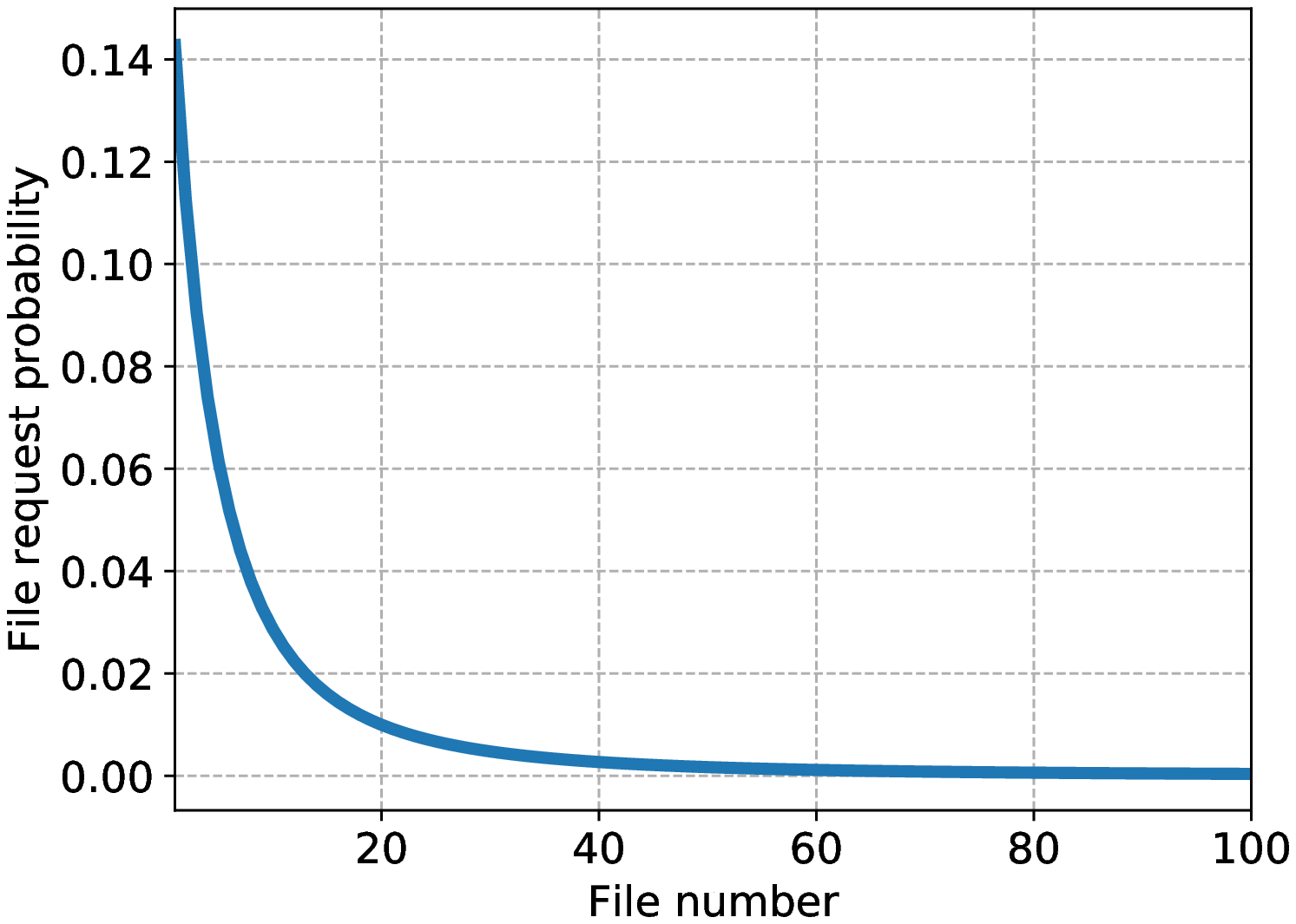}
  \caption{Initial file popularity distribution}
  \label{fig:file_pop_dist}
  \vspace{-4mm}
\end{figure}

We create a discrete-event simulator, written in Python to assess the performance of caching and coding at the wireless edge station. We consider a WiFi network consisting of an access point (AP), 10 wireless clients and a video content server consisting of 100 video files. An 802.11n AP is considered, and the data link multicast rate is set at 24 Mbps. Also, we consider that there is no delay in fetching a video segment from the cache. The video files used in the simulation are short video clips of length 2-5 minutes. The segments are encoded to a bitrate of 5 Mbps, 1280x720 resolution, and 4 seconds segment playback duration using MP4Box \cite{mp4box} DASH encoder. The segment size distribution is shown in Fig. \ref{fig:segment_dist}, where the histogram shows the fraction of different segment sizes, and the red line shows the trend. We use the MZipf file popularity distribution described in Sec. \ref{subsec:file_pop_model} with parameters $q=10$ and $\gamma=2.5$, as shown in Fig. \ref{fig:file_pop_dist} for the intial file popularity distribution. The 20 most popular files constitute approximately 80\% of the file request probability. We consider a setting where the clients are actively requesting video files for 3 hours with the mean waiting time, $1 \backslash \lambda_{S}=5$ seconds.

In order to perform the simulations, first, we generate the request profiles for different values of $M=0.5,0.10,0.15$ and $\alpha=0, 0.25, 0.5, 0.75, 1.0$, where the values of $M$ are represented in terms of the fraction of the total file size present at the video streaming server. A request profile is a sequence of user and file request pairs, which indicate the order in which the clients request the video files. Using the request profiles generated, we run simulations of the proposed system described in Sec. \ref{sec:proposed} for the cases when index coding is enabled, and when index coding is disabled. 

\section{Performance metrics and Results}
\label{sec:metrics}
We consider the following metrics to determine the performance of caching and index coding, as well as the tradeoffs between them.

\begin{figure*}
  \centering
  \subfloat[$M=0.05$]{\includegraphics[width=0.36\textwidth]{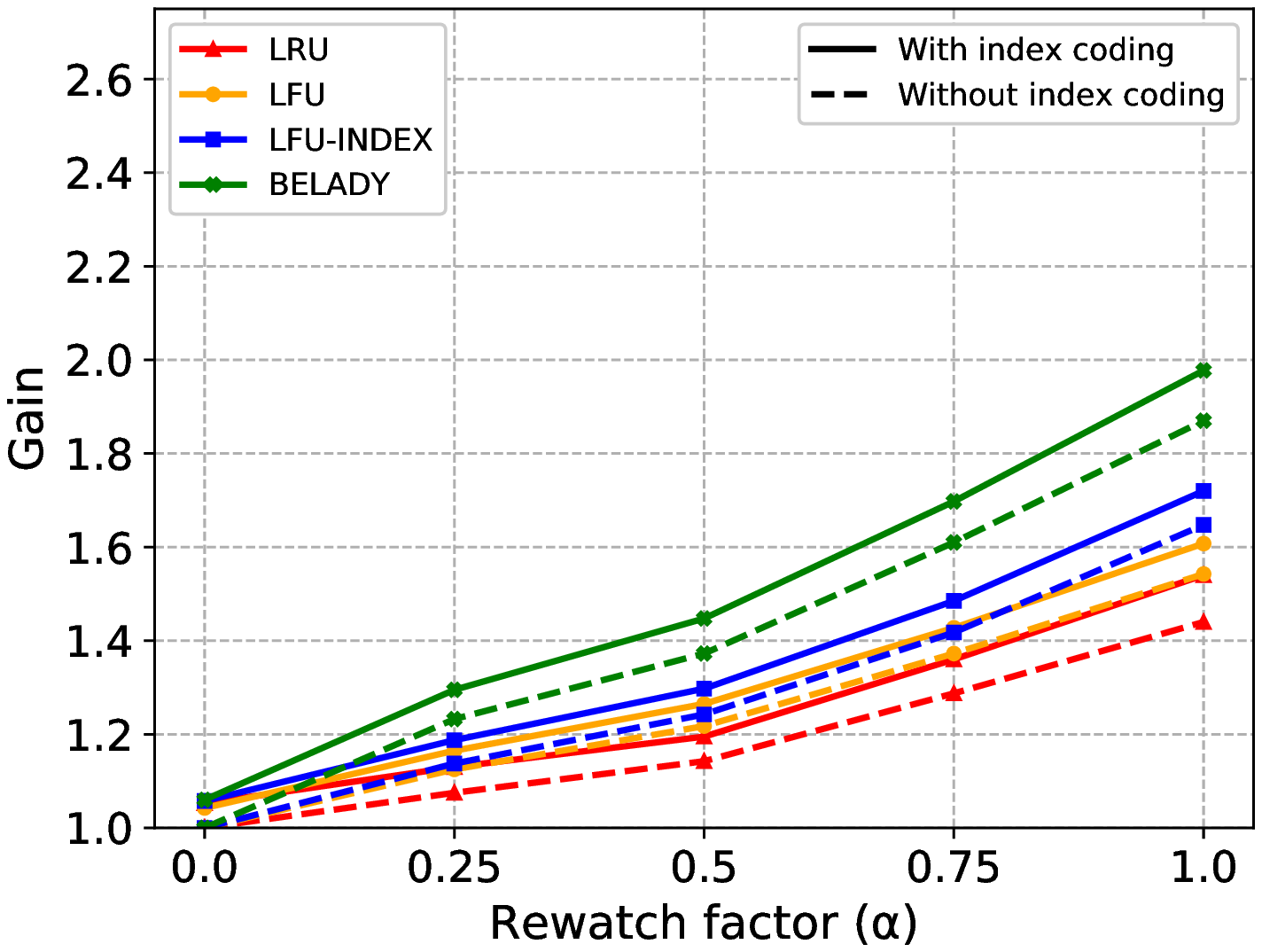}
  \hspace{-7mm}
  \label{fig:gain400} } 
  \subfloat[$M=0.10$]{\includegraphics[width=0.36\textwidth]{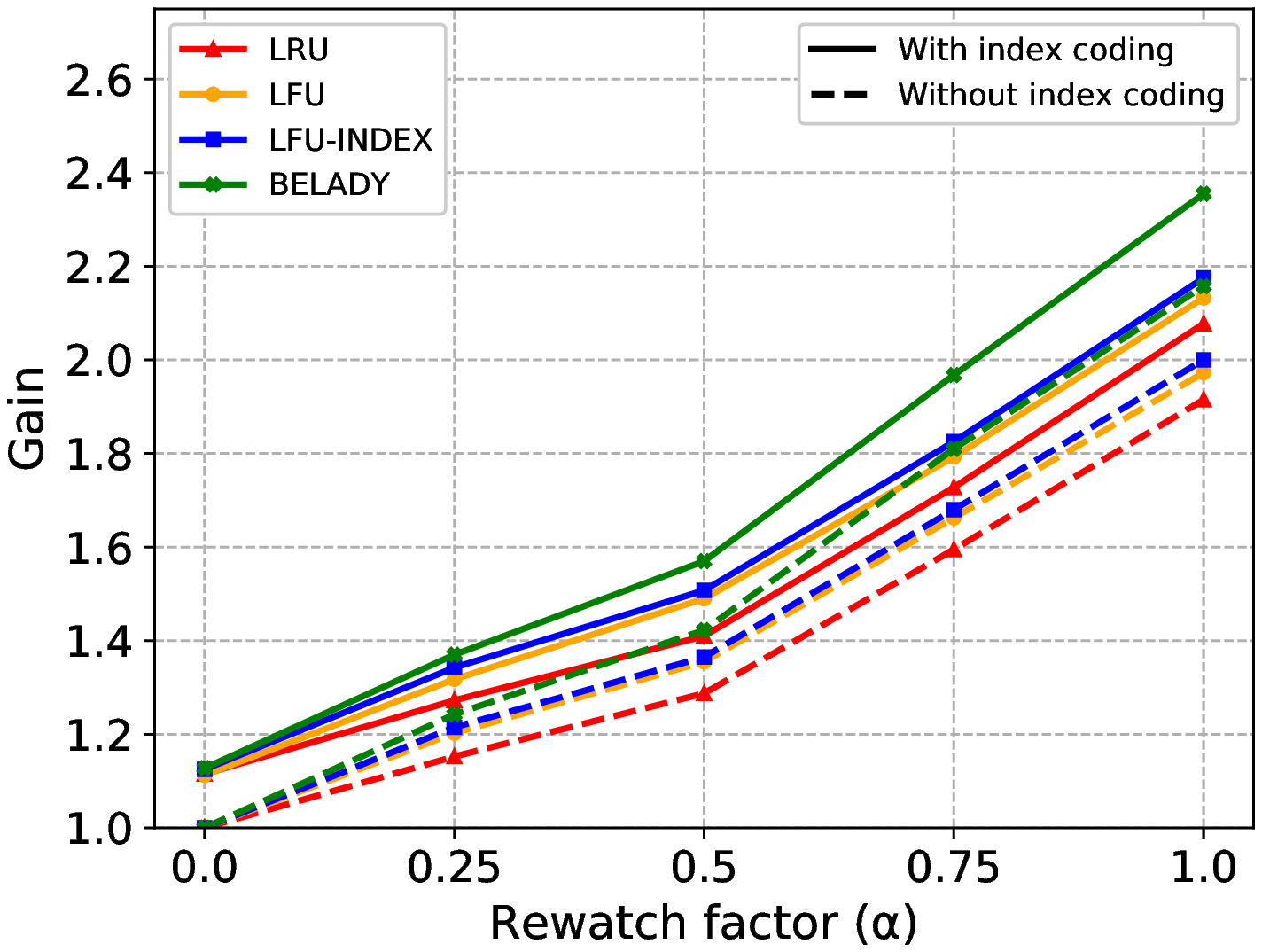}
  \hspace{-7mm}
  \label{fig:gain800} }
  \subfloat[$M=0.15$]{\includegraphics[width=0.36\textwidth]{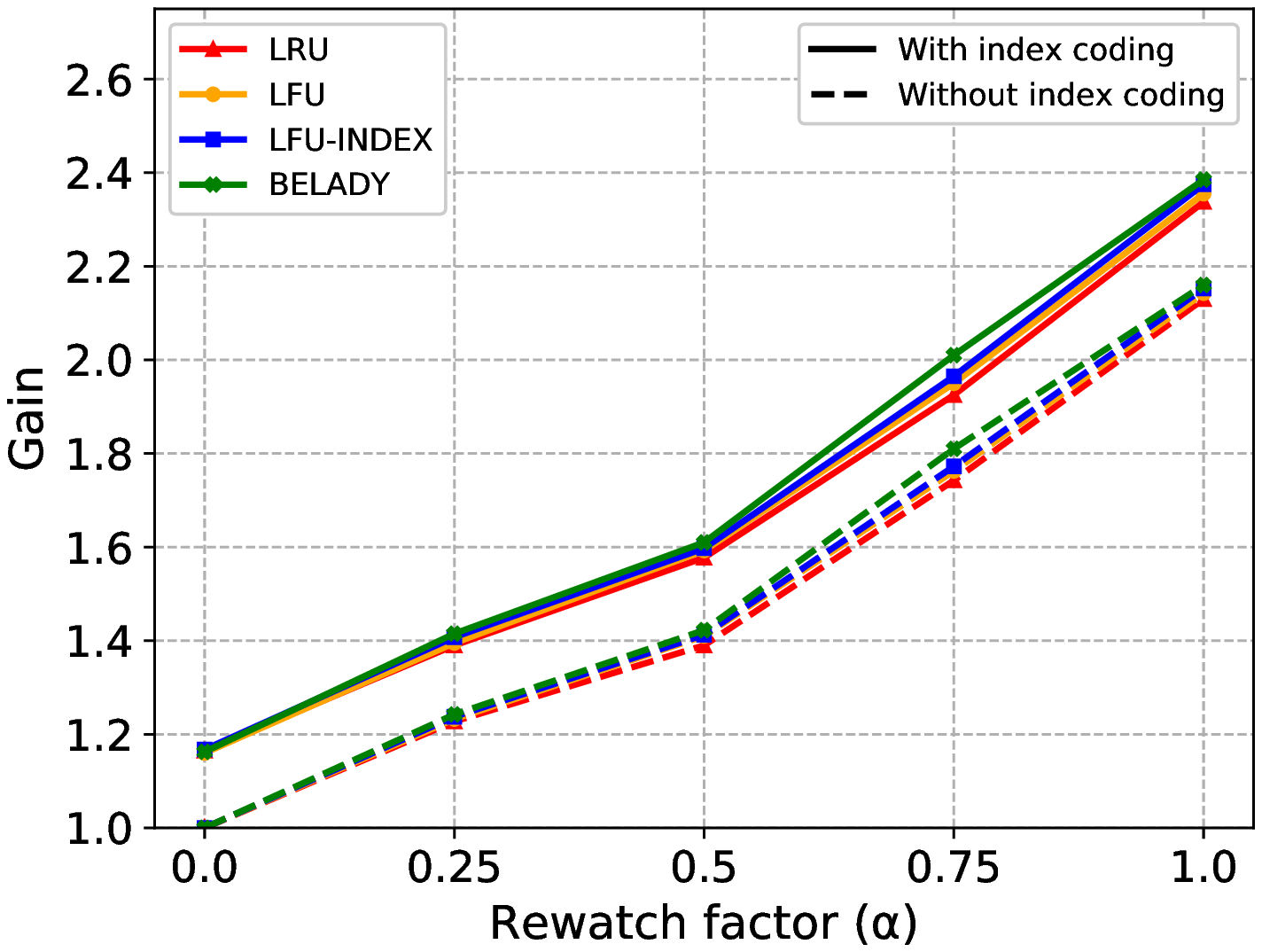} 
  \label{fig:gain1200} }
  \caption{Gain as a function of rewatch factor ($\alpha$) for different values of cache size ($M$)}
  \label{fig:gain_all}
  
  \vspace{-7mm}
\end{figure*}

\begin{figure*}
  \centering
  \subfloat[$M=0.05$]{\includegraphics[width=0.36\textwidth]{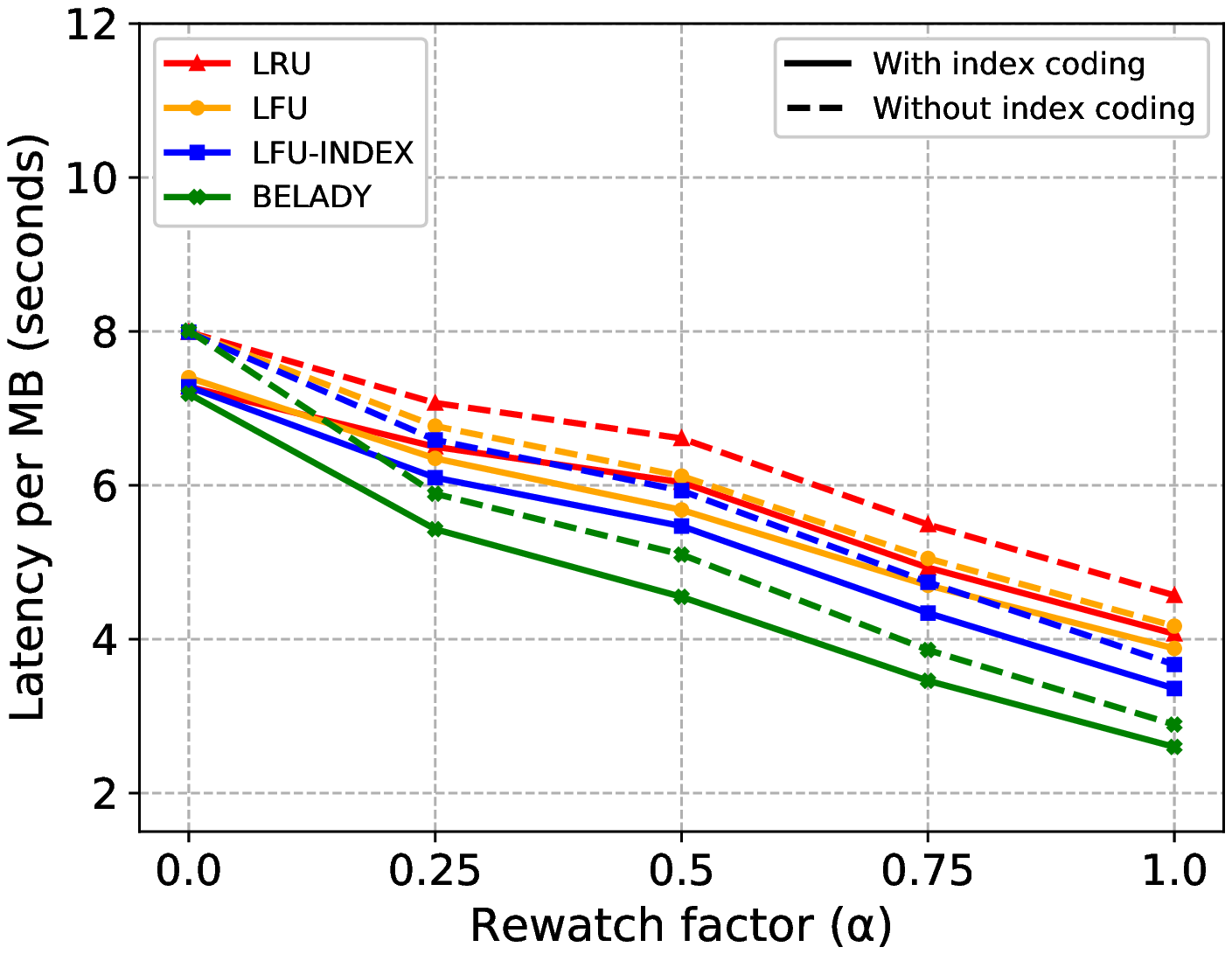} 
  \hspace{-7mm}
  \label{fig:latency400} } 
  \subfloat[$M=0.10$]{\includegraphics[width=0.36\textwidth]{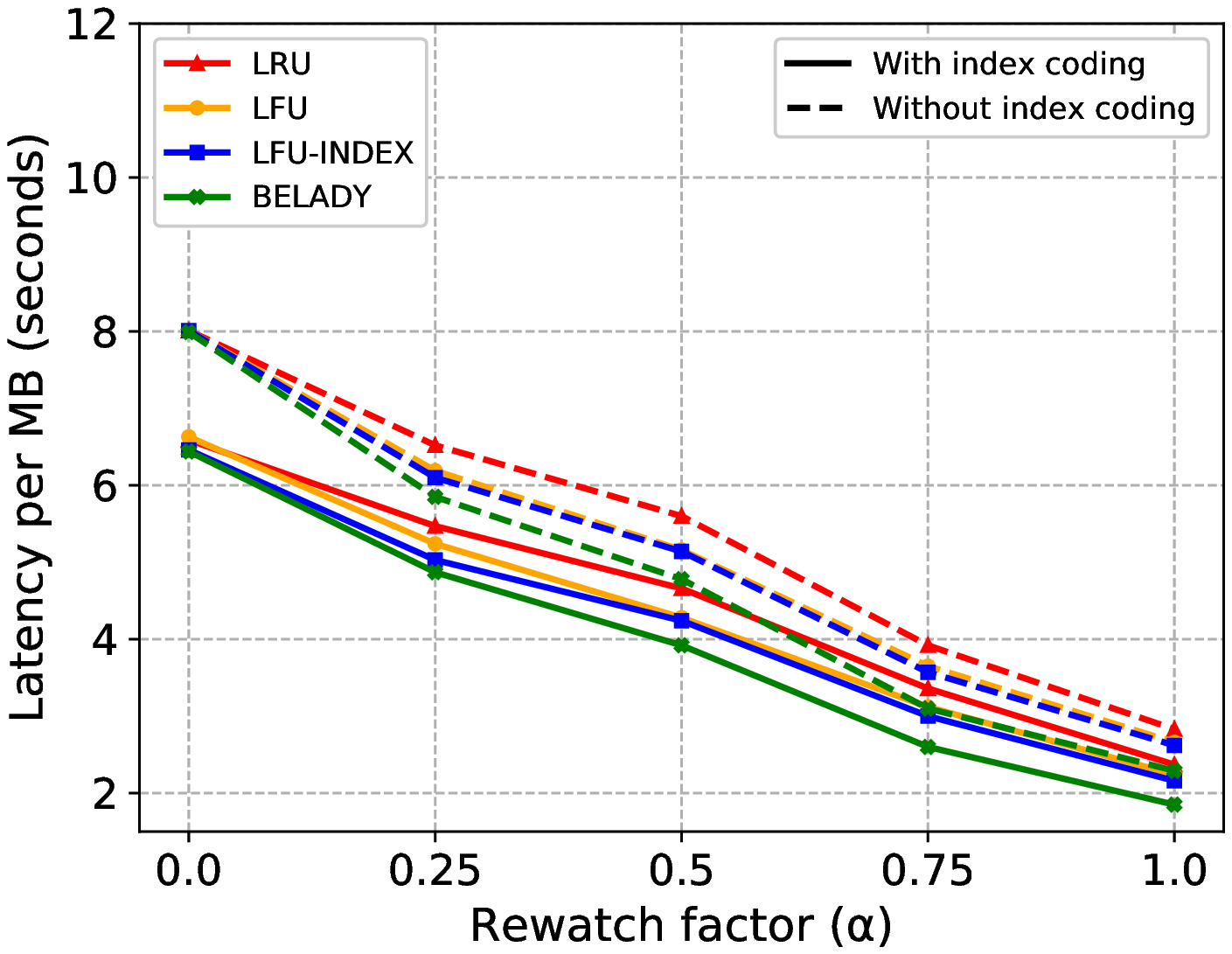} 
  \hspace{-7mm}
  \label{fig:latency800} }
  \subfloat[$M=0.15$]{\includegraphics[width=0.36\textwidth]{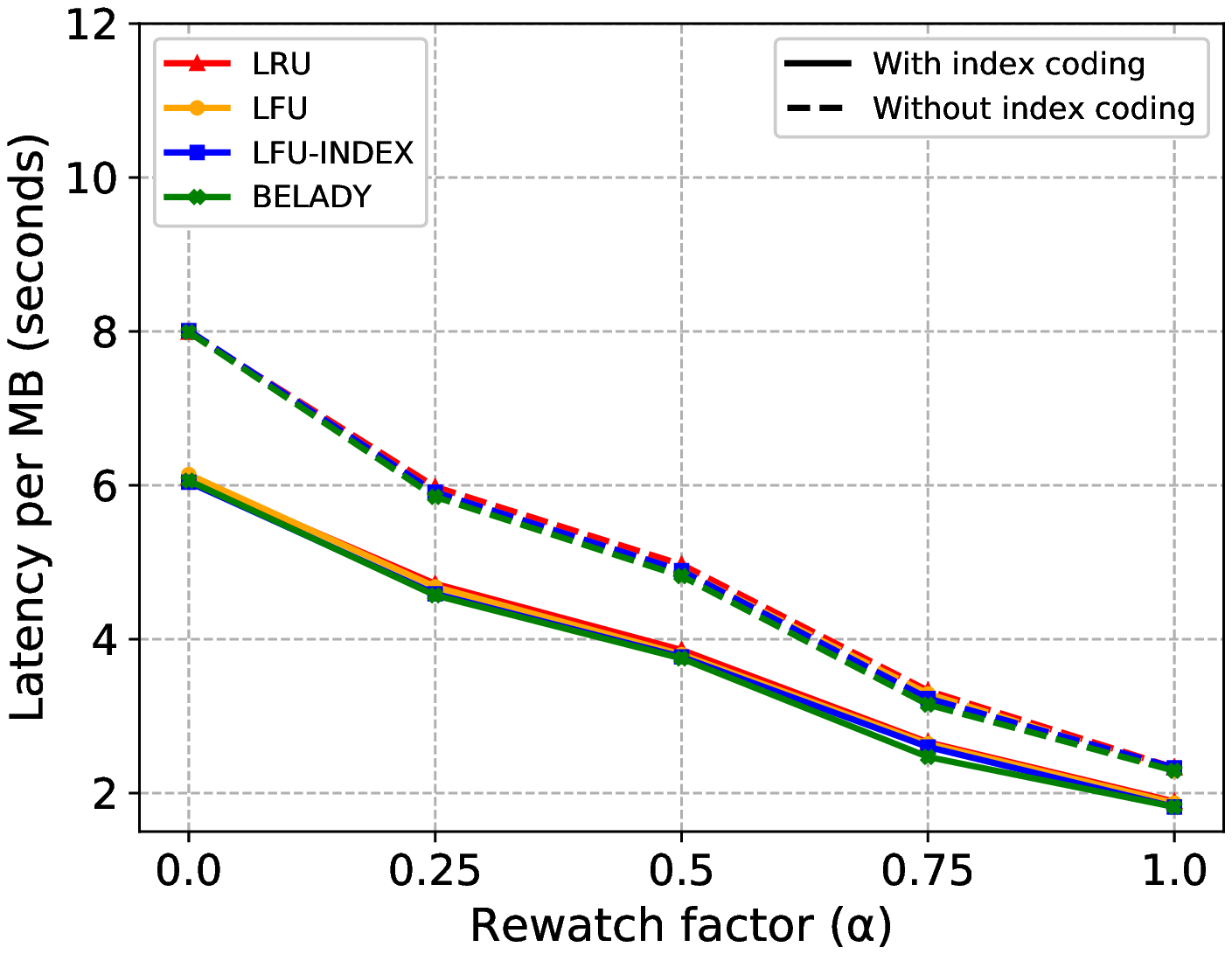} 
  \hspace{-7mm}
  \label{fig:latency1200} }
  \caption{Average perceived latency per MB as a function of rewatch fator ($\alpha$) for different values of cache size ($M$)}
  \label{fig:latency_all}
   \vspace{-5mm}
\end{figure*}

We define the {\em gain due to caching only}, $G_{c}$, as the ratio of the data transmitted in bits when caching is not employed at the client side to the data transmitted in bits when caching is employed at the client side, i.e.,
\[
G_{c} = \frac{\text{TX bits with no cache}}{\text{TX bits with cache}}
\]

We define the {\em gain due to index coding only}, $G_{i}$, as the ratio of the data transmitted in bits when only caching is employed at the client side to the data transmitted in bits when caching and index coding are employed at the client side, i.e.,
\[
\begin{aligned}
G_{i} &= \frac{\text{TX bits with cache and no index coding}}{\text{TX bits with cache and index coding}}
\end{aligned}
\]

We define the {gain due to caching and index coding}, $G_{c,i}$, as the ratio of the data transmitted in bits when no caching and index coding are employed to the data transmitted in bits when caching and index coding are employed at the client side, i.e.,
\[
\begin{aligned}
G_{c,i} &= \frac{\text{TX bits  no cache and no index coding}}{\text{TX bits  cache and index coding}} = G_{c} \times G_{i}
\end{aligned}
\]

\begin{figure*}
  \centering
  \subfloat[$M=0.05$]{\includegraphics[width=0.36\textwidth]{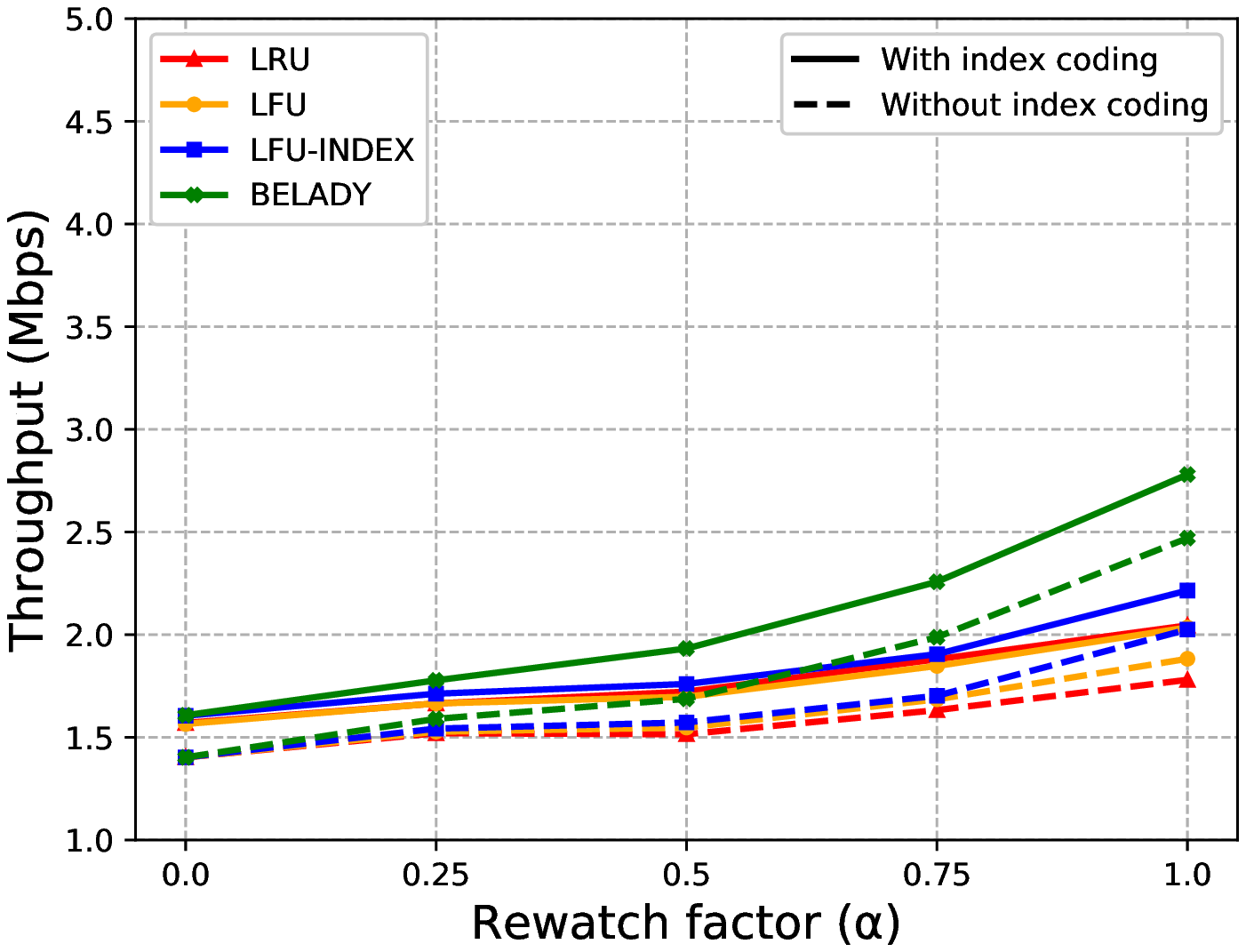} 
  \hspace{-7mm}
  \label{fig:throughput400} } 
  \subfloat[$M=0.10$]{\includegraphics[width=0.36\textwidth]{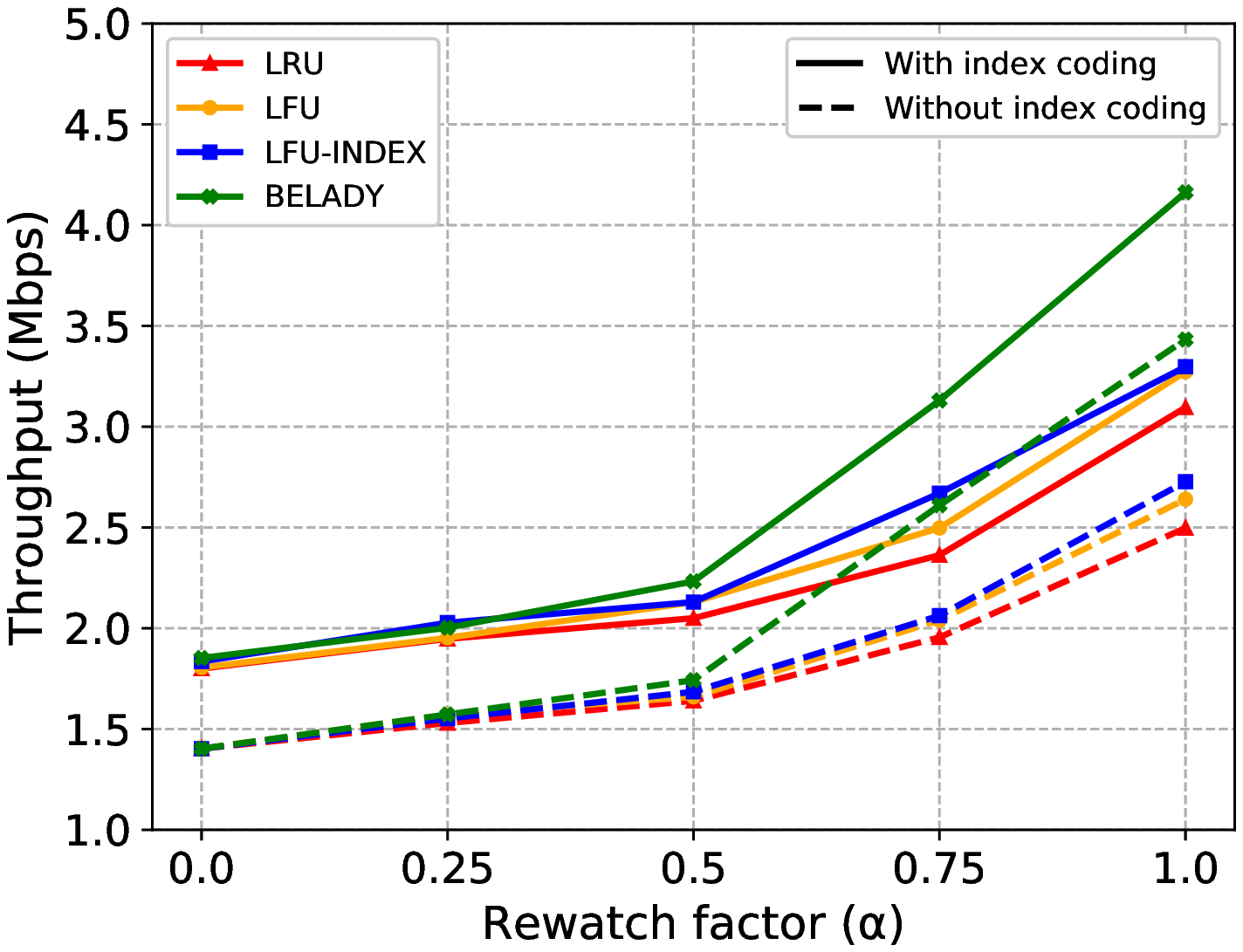} 
  \hspace{-7mm}
  \label{fig:throughput800} }
  \subfloat[$M=0.15$]{\includegraphics[width=0.36\textwidth]{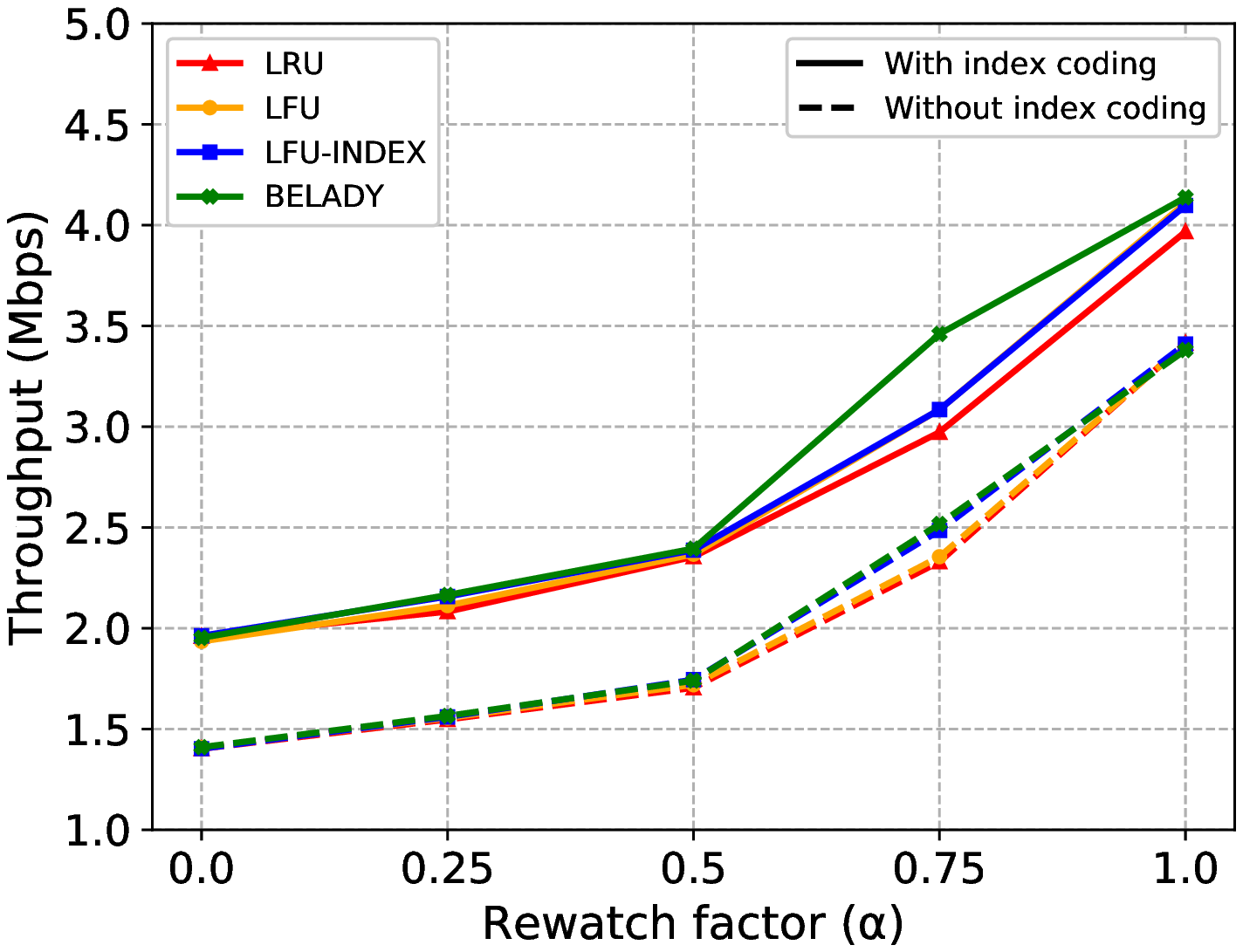} \hspace{-7mm}
  \label{fig:throughput1200} }
  \caption{Average perceived network throughput as a function of rewatch factor ($\alpha$) for different values of cache size ($M$)}
  \label{fig:throughput_all}
  \vspace{-2mm}
\end{figure*}

We define the {\em average perceived latency per MB}, $L_{s}$, as the average time elapsed between when a segment of size 1 megabyte (MB) is requested by a client, and when the segment is available for playback at the client. The perceived latency per MB is calculated for all the segments, i.e., segments which are obtained from the network and the cache.

We define the {\em average perceived throughput}, $T_{s}$, as the average number of bits received per second from the wireless edge station for every segment request made by a client. The 
perceived throughput is calculated only for the segments which are obtained from the network.

Figs. \ref{fig:gain_all}, \ref{fig:latency_all}, and \ref{fig:throughput_all} show the performance of the proposed system in terms of gain, latency, and throughput for different values of rewatch factor ($\alpha$) and cache size ($M$) respectively.
From these figures, we can see that as $\alpha$ increases, there is an improvement in gain, latency, and throughput for all the cache replacement policies described in Sec. \ref{subsec:cache_model}. The improvement in performance is due to rewatching of videos, which increases as $\alpha$ increases.

Now, consider the case where $\alpha=0$, i.e., when the clients do not rewatch videos. The improvement in performance is only due to index coding, and it also increases as the cache size ($M$) at the client increases. This is due to the increase in index coding opportunities as $M$ increases.

Moreover, for any $\alpha$, the performance due to both caching and index coding increases with increase in $M$ due to the increase in index coding and cache hit as the number of video segments stored at the cache increases. However, the improvement in gains due to caching is slightly higher than the improvement in index coding gains with increase in $M$.

Further, from the figures, we can observe that the proposed LFU-Index cache replacement policy performs better than the traditional policies---LRU and LFU, and it is closer to Belady, which is the optimal cache replacement policy for a given request profile. The gain in performance by LFU-Index is due to the improvement in index coding opportunities in addition to the cache hit count. In general, index coding opportunities arise when the cache content of the clients are different, and each client request contents present at the other clients' caches. The proposed LFU-Index policy maximizes the difference in the cache content across the clients, which significantly improves the index coding opportunities compared to the previous policies. Finally, we can see that for very large values of $M$, the performance of LFU, LRU, and LFU-Index approaches the optimum Belady as the most of the requests are served by the caches.

\section{Conclusions}
\label{sec:conclusions}
In this paper, we proposed a system model to evaluate the gains of caching and index coding, as well as the tradeoffs between them. We considered the case where an online caching policy is employed at the user to generate the side information for index coding. We also proposed a index coding heuristic at the wireless edge station, and a cache replacement policy at the user for the proposed system model. Through an extensive simulation study, we presented the effect of rewatch factor ($\alpha$) and cache size ($M$) on the performance of the system for the cases where index coding is enabled and disabled. We also showed that the LFU-Index cache replacement policy that we propose performs better than the traditional LRU and LFU policies, because it improves both the index coding opportunities and the cache hits. The LFU-index policy is being built into the Wi-Cache system that is described in \cite{wicache, wicache2}

\bibliographystyle{IEEEtran}
\bibliography{main}

\begin{thebibliography}{10}
\providecommand{\url}[1]{#1}
\csname url@samestyle\endcsname
\providecommand{\newblock}{\relax}
\providecommand{\bibinfo}[2]{#2}
\providecommand{\BIBentrySTDinterwordspacing}{\spaceskip=0pt\relax}
\providecommand{\BIBentryALTinterwordstretchfactor}{4}
\providecommand{\BIBentryALTinterwordspacing}{\spaceskip=\fontdimen2\font plus
\BIBentryALTinterwordstretchfactor\fontdimen3\font minus
  \fontdimen4\font\relax}
\providecommand{\BIBforeignlanguage}[2]{{%
\expandafter\ifx\csname l@#1\endcsname\relax
\typeout{** WARNING: IEEEtran.bst: No hyphenation pattern has been}%
\typeout{** loaded for the language `#1'. Using the pattern for}%
\typeout{** the default language instead.}%
\else
\language=\csname l@#1\endcsname
\fi
#2}}
\providecommand{\BIBdecl}{\relax}
\BIBdecl

\bibitem{cisco}
\BIBentryALTinterwordspacing
Cisco, ``{Cisco visual networking index: Global mobile data traffic forecast
  update, 2015-2020},'' White Paper, 2015. [Online]. Available:
  \url{http://goo.gl/tZ6QMk}
\BIBentrySTDinterwordspacing

\bibitem{femto}
N.~{Golrezaei}, A.~F. {Molisch}, A.~G. {Dimakis}, and G.~{Caire},
  ``Femtocaching and device-to-device collaboration: A new architecture for
  wireless video distribution,'' \emph{IEEE Commun. Mag.}, vol.~51, no.~4, pp.
  142--149, April 2013.

\bibitem{livingedge}
E.~{Bastug}, M.~{Bennis}, and M.~{Debbah}, ``Living on the edge: The role of
  proactive caching in 5g wireless networks,'' \emph{IEEE Commun. Mag.},
  vol.~52, no.~8, pp. 82--89, Aug. 2014.

\bibitem{wirelessdtod}
M.~{Ji}, G.~{Caire}, and A.~F. {Molisch}, ``Wireless device-to-device caching
  networks: Basic principles and system performance,'' \emph{IEEE J. Sel. Areas
  Commun.}, vol.~34, no.~1, pp. 176--189, Jan. 2016.

\bibitem{energy}
D.~{Liu} and C.~{Yang}, ``Energy efficiency of downlink networks with caching
  at base stations,'' \emph{IEEE J. Sel. Areas Commun.}, vol.~34, no.~4, pp.
  907--922, April 2016.

\bibitem{pushed}
K.~{Wang}, Z.~{Chen}, and H.~{Liu}, ``Push-based wireless converged networks
  for massive multimedia content delivery,'' \emph{IEEE Trans. Wireless
  Commun.}, vol.~13, no.~5, pp. 2894--2905, May 2014.

\bibitem{vidaware}
H.~{Ahlehagh} and S.~{Dey}, ``Video-aware scheduling and caching in the radio
  access network,'' \emph{IEEE/ACM Trans. Netw.}, vol.~22, no.~5, pp.
  1444--1462, Oct. 2014.

\bibitem{informed}
B.~D. Higgins, J.~Flinn, T.~J. Giuli, B.~Noble, C.~Peplin, and D.~Watson,
  ``Informed mobile prefetching,'' in \emph{Proc. 2012 ACM MobiSys}, 2012, pp.
  155--168.

\bibitem{onlinecoded}
R.~{Pedarsani}, M.~A. {Maddah-Ali}, and U.~{Niesen}, ``Online coded caching,''
  \emph{IEEE/ACM Trans. Netw.}, vol.~24, no.~2, pp. 836--845, April 2016.

\bibitem{fundamental}
M.~A. Maddah-Ali and U.~Niesen, ``{Fundamental limits of caching},'' \emph{IEEE
  Trans. Inf. Theory}, vol.~60, no.~5, pp. 2856--2867, May 2014.

\bibitem{decentralized}
M.~A. Maddah{-}Ali and U.~Niesen, ``{Decentralized coded caching attains
  order-optimal memory-rate tradeoff},'' \emph{IEEE/ACM Trans. Netw.}, vol.~23,
  no.~4, pp. 1029--1040, Aug 2015.

\bibitem{iscod}
Y.~Birk and T.~Kol, ``{Coding on demand by an informed source (ISCOD) for
  efficient broadcast of different supplemental data to caching clients},''
  \emph{IEEE Trans. Inf. Theory}, vol.~52, no.~6, pp. 2825--2830, June 2006.

\bibitem{onlinerandom}
Q.~{Yan}, U.~{Parampalli}, X.~{Tang}, and Q.~{Chen}, ``Online coded caching
  with random access,'' \emph{IEEE Commun. Lett.}, vol.~21, no.~3, pp.
  552--555, March 2017.

\bibitem{wicache}
L.~{Chhangte}, A.~{Garg}, D.~{Manjunath}, and N.~{Karamchandani}, ``Wi-cache:
  Towards an sdn based distributed content caching system in wlan,'' in
  \emph{Proc. 2018 COMSNETS}, Jan 2018, pp. 503--506.

\bibitem{wicache2}
L.~Chhangte, D.~Manjunath, and N.~Karamchandani, ``An sdn based content cache
  at the wifi edge,'' in \emph{Proc. 2018 ACM MobiCom}, Oct. 2018, pp.
  672--674.

\bibitem{bbc}
M.~{Lee}, M.~{Ji}, A.~F. {Molisch}, and N.~{Sastry}, ``Throughput–outage
  analysis and evaluation of cache-aided d2d networks with measured popularity
  distributions,'' \emph{IEEE Trans. Wireless Commun.}, vol.~18, no.~11, pp.
  5316--5332, Nov. 2019.

\bibitem{indexalgorithms}
M.~A.~R. Chaudhry and A.~Sprintson, ``{Efficient algorithms for index
  coding},'' in \emph{IEEE INFOCOM Workshops 2008}, April 2008, pp. 1--4.

\bibitem{lru}
S.~Angelopoulos, R.~Dorrigiv, and A.~L\'{o}pez-Ortiz, ``On the separation and
  equivalence of paging strategies,'' in \emph{Proc. 2007 ACM-SIAM}, 2007, pp.
  229--237.

\bibitem{lfu}
{D. Lee {\em et al.}}, ``Lrfu: a spectrum of policies that subsumes the least
  recently used and least frequently used policies,'' \emph{IEEE Trans.
  Comput.}, vol.~50, no.~12, pp. 1352--1361, Dec. 2001.

\bibitem{belady}
L.~A. {Belady}, ``A study of replacement algorithms for a virtual-storage
  computer,'' \emph{IBM Systems Journal}, vol.~5, no.~2, pp. 78--101, 1966.

\bibitem{mp4box}
``{MP4Box GPAC},'' \url{https://gpac.wp.imt.fr/mp4box/}.

\end{thebibliography}

\end{document}